\documentclass[12 pt, amsfonts, amssymb,color]{article}

\usepackage{amsmath,amssymb,amsfonts,latexsym,float,graphics,epsfig}

\begin{document}

\renewcommand\theequation{\arabic{section}.\arabic{equation}}
\catcode`@=11 \@addtoreset{equation}{section}
\newtheorem{axiom}{Definition}[section]
\newtheorem{theorem}{Theorem}[section]
\newtheorem{axiom2}{Example}[section]
\newtheorem{lem}{Lemma}[section]
\newtheorem{prop}{Proposition}[section]
\newtheorem{cor}{Corollary}[section]
\newcommand{\be}{\begin{equation}}
\newcommand{\ee}{\end{equation}}
\newcommand{\bea}{\begin{eqnarray}}
\newcommand{\eea}{\end{eqnarray}}

\newcommand{\equal}{\!\!\!&=&\!\!\!}
\newcommand{\rd}{\partial}
\newcommand{\g}{\hat {\cal G}}
\newcommand{\bo}{\bigodot}
\newcommand{\res}{\mathop{\mbox{\rm res}}}
\newcommand{\diag}{\mathop{\mbox{\rm diag}}}
\newcommand{\Tr}{\mathop{\mbox{\rm Tr}}}
\newcommand{\const}{\mbox{\rm const.}\;}
\newcommand{\cA}{{\cal A}}
\newcommand{\bA}{{\bf A}}
\newcommand{\Abar}{{\bar{A}}}
\newcommand{\cAbar}{{\bar{\cA}}}
\newcommand{\bAbar}{{\bar{\bA}}}
\newcommand{\cB}{{\cal B}}
\newcommand{\bB}{{\bf B}}
\newcommand{\Bbar}{{\bar{B}}}
\newcommand{\cBbar}{{\bar{\cB}}}
\newcommand{\bBbar}{{\bar{\bB}}}
\newcommand{\bC}{{\bf C}}
\newcommand{\cbar}{{\bar{c}}}
\newcommand{\Cbar}{{\bar{C}}}
\newcommand{\Hbar}{{\bar{H}}}
\newcommand{\cL}{{\cal L}}
\newcommand{\bL}{{\bf L}}
\newcommand{\Lbar}{{\bar{L}}}
\newcommand{\cLbar}{{\bar{\cL}}}
\newcommand{\bLbar}{{\bar{\bL}}}
\newcommand{\cM}{{\cal M}}
\newcommand{\bM}{{\bf M}}
\newcommand{\Mbar}{{\bar{M}}}
\newcommand{\cMbar}{{\bar{\cM}}}
\newcommand{\bMbar}{{\bar{\bM}}}
\newcommand{\cP}{{\cal P}}
\newcommand{\cQ}{{\cal Q}}
\newcommand{\bU}{{\bf U}}
\newcommand{\bR}{{\bf R}}
\newcommand{\cW}{{\cal W}}
\newcommand{\bW}{{\bf W}}
\newcommand{\bZ}{{\bf Z}}
\newcommand{\Wbar}{{\bar{W}}}
\newcommand{\Xbar}{{\bar{X}}}
\newcommand{\cWbar}{{\bar{\cW}}}
\newcommand{\bWbar}{{\bar{\bW}}}
\newcommand{\abar}{{\bar{a}}}
\newcommand{\nbar}{{\bar{n}}}
\newcommand{\pbar}{{\bar{p}}}
\newcommand{\tbar}{{\bar{t}}}
\newcommand{\ubar}{{\bar{u}}}
\newcommand{\utilde}{\tilde{u}}
\newcommand{\vbar}{{\bar{v}}}
\newcommand{\wbar}{{\bar{w}}}
\newcommand{\phibar}{{\bar{\phi}}}
\newcommand{\Psibar}{{\bar{\Psi}}}
\newcommand{\bLambda}{{\bf \Lambda}}
\newcommand{\bDelta}{{\bf \Delta}}
\newcommand{\p}{\partial}
\newcommand{\om}{{\Omega \cal G}}
\newcommand{\ID}{{\mathbb{D}}}
\newcommand{\pr}{{\prime}}
\newcommand{\prr}{{\prime\prime}}
\newcommand{\prrr}{{\prime\prime\prime}}

\title{B\"acklund Transformation and Quasi-Integrable Deformation of Mixed Fermi-Pasta-Ulam and Frenkel-Kontorova Models}
\author{Kumar Abhinav$^a$\footnote{E-mail: kumar.abhinav@bose.res.in}, A Ghose Choudhury$^b$\footnote{E-mail aghosechoudhury@gmail.com} and Partha Guha$^a$\footnote{E-mail: partha@bose.res.in}
\and
$^a$SN Bose National Centre for Basic Sciences \\
JD Block, Sector III, Salt Lake, Kolkata 700106,  India\\
$^b$Department of Physics, Surendranath  College,\\ 24/2 Mahatma Gandhi Road, Calcutta 700009, India\\}

\date{ }

\maketitle

\smallskip
\begin{center}
{\it Dedicated to the memory of Anjan Kundu}
\end{center}

\bigskip

\begin{abstract}
In this paper we study a non-linear partial differential equation (PDE), proposed by Kudryashov [arXiv:1611.06813v1[nlin.SI]], 
using continuum limit approximation of mixed Fermi-Pasta-Ulam and Frenkel-Kontorova Models. 
This generalized semi-discrete equation can be considered as a model for the description of non-linear dislocation 
waves in crystal lattice and the corresponding continuous system can be called mixed generalized potential KdV and
sine-Gordon equation. We obtain the B\"acklund transformation of
this equation in Riccati form in inverse method. We further study the quasi-integrable deformation of this model.
\end{abstract}

\smallskip


\smallskip


\section{Introduction}
Recently  Kudryashov \cite{Kud1}, using the continuous limit approximation, has derived a 
nonlinear partial differential equation 
for the description of dislocations in a crystalline lattice which can be regarded as a generalization both the Frenkel-Kontorova 
\cite{KF,BK} and 
the Fermi-Pasta-Ulam \cite{FPU,FPU1,ZK} models. This  generalized  model  can  be  considered
to describe nonlinear dislocation waves in the crystal lattice following,

\bea
&&\frac{d^2 y_i}{dt^2} = (y_{i+1} − 2y_i + y_{i−1} ) \Big[k + \alpha (y_{i+1} − y_{i−1} ) + \beta (y_{i+1}^{2} + y_{i}^{2} + y_{i−1}^{2} - y_{i+1}y_{i} - y_{i+1}y_{i−1} - y_iy_{i−1} )\Big]\nonumber\\
&&\qquad\qquad - f_0 \sin\big(\frac{2\pi y_i}{a} \big)
\eea
for all $(i = 1, . . . , N)$. Here $y_i$ denotes the displacement of the $i$-th mass from its original position,
$t$ is time, $k, \alpha, \beta, f_0$ and $a$ are constant parameters of the system. This equation boils down to
Frenkel-Kontorova form for the description of
dislocations in the rigid body when $\alpha = 0$ and $\beta = 0$. In the case of $f = 0$ and $\beta = 0$ the system of equations 
becomes 
the well-known Fermi-Pasta-Ulam model. One must note that the Fermi-Pasta-Ulam model $N \to \infty$ and $h \to 0$ is transformed
to the Korteweg-de Vries (KdV) equation.

\paragraph*{}Kudryashov showed that in the continuum limit the semi-discrete equation takes the following form:
\be\label{E1}u_{xt}+\alpha u_x u_{xx}+3\beta u_x^2u_{xx}+\gamma u_{xxxx}=\delta \sin u.\ee
In the special case when $\alpha=0$ and $\gamma=2\beta$ it is shown that one can cast the equation in the AKNS scheme with the Lax pair given by
\be L=\left(\begin{array}{cccc} -i\lambda & q\\ r & i\lambda\end{array}\right), \;\;\;M=\left(\begin{array}{cccc}
A & B\\ C & -A\end{array}\right),\label{N01}\ee
where $q=-r=-u_x/2$ and the functions $A$, $B$ and $C$ are
$$A=-\frac{\delta}{4}\cos u (i\lambda)^{-1}+\beta u_x^2(i\lambda) +8\beta(i\lambda)^3,$$
$$B=-\frac{\delta}{4}\sin u (i\lambda)^{-1} + \left(\beta u_{xxx}+\frac{\beta}{2}u_x^3\right)-2\beta u_{xx}(i\lambda) +4\beta u_x (i\lambda)^2,$$
$$C=-\frac{\delta}{4}\sin u (i\lambda)^{-1} - \left(\beta u_{xxx}+\frac{\beta}{2}u_x^3\right)-2\beta u_{xx}(i\lambda) -4\beta u_x (i\lambda)^2.$$
Equation (\ref{E1}) follows from the usual zero-curvature condition, viz
\be\label{E3} F_{tx}=L_t-M_x+[L , M]=0.\ee
As (\ref{E1}) is a continuous system one may derive B\"{a}cklund transformations \cite{FT,Das} for the equation using the
method proposed by Konno and Wadati \cite{KW,KW1} which makes use of the  Riccati equation. In the next section we study
this procedure. 

\paragraph*{}Real physical systems contain finite number of degrees of freedom, thereby prohibiting the corresponding field-theoretical
models from being integrable in principle, as the latter requires infinite conserved quantities. However, they are known to posses 
physically obtainable solitonic states, considerably
similar in structure to integrable ones, like sine-Gordon (SG) \cite{FerrSG}. Therefore, the study of physical continuous
systems can be motivated as slightly deformed integrable models. Recently \cite{FZ,FZ1}, the SG model was
deformed into an approximate system, with a {\it finite} number of conserved charges. The corresponding
connection (curvature) was almost flat, yielding an anomaly instead of the usual zero-curvature condition. This system was,
therefore, dubbed as quasi-integrable
(QI). In a recent paper \cite{N0} we obtain the quasi-integrable deformation of the KdV equation. In the present paper we outline
the quasi-deformation of the new equation proposed by Kudryashov.

\subsection{Lagrangian and Hamiltonian}\label{1.1}
The integrable sector of Eq. \ref{E1}, represented as,

\be
u_{xt}+3\beta u_x^2u_{xx}+2\beta u_{xxxx}=\delta \sin u,\label{N1}
\ee
follows from a Euler-Lagrange form,

\be
\left(\frac{\partial{\cal L}}{\partial u_t}\right)_t+\left(\frac{\partial{\cal L}}{\partial u_x}\right)_x+\left(\frac{\partial{\cal L}}{\partial u_{xxx}}\right)_{xxx}=\frac{\partial{\cal L}}{\partial u},\label{N2}
\ee
corresponding the Lagrangian,

\be
{\cal L}=\frac{1}{2}u_xu_t+\frac{\beta}{4}u_x^4+\beta u_xu_{xxx}-\delta\cos u.\label{N3}
\ee
Therefore, the system described by Eq. \ref{N1} has a Lagrangian structure with respect to the variable $u$. However, the
presence of higher order derivatives therein depicts time-evolution of $u_x$ instead. However, the canonical variable is
identified to be $u$, leading to a Hamiltonian through the Legendre transformation,

\be
{\cal H}:=\frac{\partial {\cal L}}{\partial u_t}-{\cal L}\equiv \delta\cos u-\frac{\beta}{4}u_x^4-\beta u_xu_{xxx}.\label{N02}
\ee

\section{B\"{a}cklund transformation using the Riccati equation}
Classical stable solutions of integrable systems are of paramount interest owing to their Physical realizability. The
standard way to arrive at such a solution is through B\"acklund transformation \cite{Das}. For this purpose, we write
the scattering problem as,
\be\label{E5}\frac{\partial}{\partial x}\left(\begin{array}{ccc} \psi_1\\ \psi_2\end{array}\right)=
\left(\begin{array}{cccc} -i\lambda & q\\ r & i\lambda\end{array}\right)\left(\begin{array}{ccc} \psi_1\\ \psi_2\end{array}\right),\ee
\be\label{E7}\frac{\partial}{\partial t}\left(\begin{array}{ccc} \psi_1\\ \psi_2\end{array}\right)=
\left(\begin{array}{cccc} A & B\\ C & -A\end{array}\right)\left(\begin{array}{ccc} \psi_1\\ \psi_2\end{array}\right).\ee
The consistency of (\ref{E5}) and (\ref{E7}) yields the equation (\ref{E3}) with the eigenvalues $\lambda$ being time independent. By introducing the function
\be \Gamma:=\frac{\psi_1}{\psi_2}\ee we obtain from (\ref{E5}) and (\ref{E7}) the Riccati equations:
\be\label{E9}\frac{\partial \Gamma}{\partial x}=-2i\lambda \Gamma +q-r\Gamma^2,\ee
\be\label{E11}\frac{\partial \Gamma}{\partial t}=B + 2A\Gamma - C\Gamma^2.\ee
To derive a B\"{a}cklund transformation one seeks a transformation $\Gamma\longrightarrow\Gamma^\prime$ which satisfies an equation identical to (\ref{E9}) with $q(x,t)$ replaced by
\be\label{E13}q^\prime(x, t)=q(x, t)+f(\lambda, \Gamma)\ee for some suitable function $f$. Then upon eliminating $\Gamma$ between (\ref{E9}) and (\ref{E13}) one arrives at the desired B\"{a}cklund transformation. In our case as $r(x,t)=-q(x,t)=u_x/2$ we shall take
\be\label{E15} \Gamma^\prime=\frac{1}{\Gamma}\;\;\;\mbox{and}\;\;\;q^\prime(x, t)=q(x, t)-2\frac{\partial}{\partial x}\tan^{-1}\Gamma.\ee
As,
$$0=\frac{\partial}{\partial x}(\Gamma\Gamma^\prime)=-4i\lambda+(q+q^\prime)(\Gamma+\Gamma^\prime)$$ (where use has been made of (\ref{E9}) and its corresponding similar version with $q^\prime$ and $\Gamma^\prime$) we see that
\be\label{E17} \frac{q+q^\prime}{2i\lambda}=\frac{2\Gamma}{1+\Gamma^2}, \;\;\;q-q^\prime=2\frac{\partial}{\partial x}\tan^{-1}\Gamma.\ee
From the latter equation we notice that as $r(x,t)=-q(x,t)=u_x/2$,
$$\tan^{-1}\Gamma=\frac{1}{4}(u^\prime-u), \;\;\;\mbox{or}\;\;\Gamma=\tan\left(\frac{u^\prime-u}{4}\right).$$
On the other hand the first part of (\ref{E17}) with $\Gamma$ as given above yields,
$$\frac{q+q^\prime}{2i\lambda}=\frac{2\Gamma}{1+\Gamma^2}=\sin\left(\frac{u^\prime-u}{2}\right),$$ whence,
\be\label{E21}u_x+u^\prime_x=4i\lambda\sin\left(\frac{u^\prime-u}{2}\right).\ee
To find the time-part we use the form of $\Gamma$ found above in (\ref{E11}) to obtain after simplification
\be \label{E23}u_t-u^\prime_t=2(C-B)+4A\sin\left(\frac{u^\prime-u}{2}\right)-2(C+B)\cos\left(\frac{u^\prime-u}{2}\right).\ee
Equations (\ref{E21}) and (\ref{E23}) constitute the desired B\"{a}cklund transformation.

\paragraph*{One-Soliton Solution:}In order to construct a { \it one-soliton solution} of (\ref{E1}) from the B\"acklund transformation 
given by (\ref{E21}) and (\ref{E23}) we set $\lambda = i\mu$, \, $\mu \in {\Bbb R}$
and note that as $u^{\prime}= 0$ is a trivial solution of (\ref{E1}) we have 
\be
u_x = 4\mu \sin\,\frac{u}{2}, \qquad    u_t = -4A \sin\,\frac{u}{2} \qquad \hbox{with} \qquad
A = \frac{\delta}{4\mu} - 8\beta\mu^3.
\ee 
 It now follows that $$ u = \pm 4 \arctan\,(e^{\theta}), \qquad \hbox{ where }\qquad \theta = 2\mu x - 
\frac{\delta t}{2\mu} + 16\beta\mu^3 t
$$
which matches the solution given in \cite{Kud1} when $\eta = 2\mu$, $\delta \to -\delta$ and $\beta \to - \beta$.

\section{Derivation of the conservation laws}
Given the Lax pair one can quite easily derive an infinite number of conserved quantities  for equation (\ref{E1}). Firstly it follows from the zero-curvature condition that
\begin{eqnarray}\label{E25}
A_x+rB-qC=0\\
q_t-B_x-2(i\lambda)B-2qA=0\\
r_t-C_x+2(i\lambda)C+2rA=0.\end{eqnarray}
Using (\ref{E11}) and the above set of equations one easily derives the following equation
\be\label{E27}
\frac{\partial}{\partial t}(r\Gamma)=\frac{\partial}{\partial x}(-A+C\Gamma),\ee
which has the general form of an conservation law for the conserved densities and the flows. In fact from (\ref{E9}) we have
\be\label{E29}r[(r\Gamma)/r]_x=-2(i\lambda)(r\Gamma)+q r-(r\Gamma)^2,\ee where the suffix represents the usual partial differentiation with respect to $x$. We may expand $(r\Gamma)$ in inverse powers of $(i\lambda)$ as
\be\label{E31} r\Gamma=\sum_{n=1}^\infty f_n (i\lambda)^{-n},\ee which when inserted into (\ref{E29}) yields the following recurrence relation for the conserved densities, namely
\be -2f_{s+1}=r\left(\frac{f_s}{r}\right)_x-qr\delta_{s, 0}+\sum_{q=1}^{n-1}f_qf_{s-q}, \;\;\;s=0,1,2,\cdots,\ee
thereby confirming integrability of the system by yielding infinite number of conserved charges. The first few these
conserved densities, at the lowest order, are given by,
\begin{eqnarray} f_1=\frac{1}{2}qr,\\
f_2=-\frac{1}{2^2}rq_x,\\
f_3=\frac{1}{2^3}rq_{xx}-\frac{1}{2^3}(qr)^2,\\
f_4=-\frac{1}{2^4}\left((q_{xxx}-(q^2r)_x)-2qrq_x\right).\end{eqnarray}
Consequently (\ref{E27}) becomes

\bea
&&\sum_{n=1}^\infty f_{n,t}(i\lambda)^{-n}=\frac{\partial}{\partial x}\Big[-A_{-1}(i\lambda)^{-1}-A_1(i\lambda)-A_3(i\lambda)^3\nonumber\\
&&\qquad\qquad\qquad\quad+\left(C_{-1}(i\lambda)^{-1}+C_0+C_1(i\lambda)+C_2(i\lambda)^2\right)\sum_{n=1}^\infty\left(\frac{f_n}{r}\right)(i\lambda)^{-n})\Big].\label{N4}
\eea
 where the   coefficients of the various powers of $(i\lambda)$ are as follows:
 $$A_1=-\frac{\delta}{4}\cos u,\;\;\; A_1=\beta u_x^2,\;\;\; A_3=8\beta,$$
 $$B_{-1}=C_{-1}=-\frac{\delta}{4}\sin u, \;\;\;\;B_0=-C_0=\beta u_{xxx}+\frac{\beta}{2}u_x^3,$$
 $$B_1=C_1=-2\beta u_{xx},\;\;\;B_2=-C_2=4\beta u_x$$
It can be verified that for all positive powers of $(i\lambda)$ this relation are identically satisfied, while for various negative powers of $(i\lambda)$ we have the following relations:
\be(i\lambda)^{-1}:\;\;\;\;\frac{\partial}{\partial t}\left(-\frac{1}{8}u_x^2\right)=
\frac{\partial}{\partial x}\left(\frac{\delta}{4}\cos u+\frac{\beta}{2}u_x u_{xxx}-\frac{\beta}{4}u_{xx}^2+\frac{3\beta}{16}u_x^4\right),\label{N5}
\ee
which leads to Eq. \ref{N1} itself. The next order continuity equation has the form,
\be(i\lambda)^{-2}:\;\;\;\;\frac{\partial}{\partial t}[-(u_x^2)_x]=\frac{\partial}{\partial x}\left[u_x\left(-2\delta\sin u+5\beta u_{xx}u_x^2+4\beta u_{xxxx}\right)\right],\ee
and so on.
\section{QI Deformation}
The most trivial way to QI-deform the present system is by deforming the periodic $\sin u$ function, like that in
Ref. \cite{FZ,FZ1} or any other way. Incorporating this into the time-component $M$ of the Lax pair as usual,
only the first terms of ${\cal O}\left(1/\lambda\right)$ in $B$ and $C$ changes. As only $C$ appears in the continuity
Eq. \ref{E27}, and further, as its ${\cal O}\left(1/\lambda\right)$ part does not contribute while obtaining Eq. \ref{N5}
from Eq. \ref{N4}, the ${\cal O}\left(1/\lambda\right)$ conserved quantity remains conserved even after the  deformation. However,
the higher order equations coming from Eq. \ref{N4} gets effected and hence they do not support conserved charges in
general. Thus, quasi-integrability is achieved by definition. Moreover, the deformation need {\it not} to be small or
finite, as far as Eq. \ref{N5} is concerned.

\paragraph*{}As an attempt for non-trivial QI deformation of the present system, we opt to deform the highest powered term in $u_x$ in
the temporal component of the Lax pair: $u_x^3\rightarrow u_x^{3+\epsilon}$. This exclusively changes the $C_0$ contribution
in Eq. \ref{N4}, leading to

\be
\frac{\partial}{\partial t}\left(-\frac{1}{8}u_x^2\right)=\frac{\partial}{\partial x}\left[\frac{\delta}{4}\cos u+\frac{\beta}{2}u_x u_{xxx}-\frac{\beta}{4}u_{xx}^2+\frac{\beta}{16}u_x^4\left(2u_x^\epsilon+1\right)\right],\label{N6}
\ee
that falls back to Eq. \ref{N5} for $\epsilon\rightarrow 0$. Here also, even finite values of $\epsilon$ will work for
conservation of the ${\cal O}\left(1/\lambda\right)$ charge. In fact, as the RHS of Eq. \ref{E27} is a total derivative,
as long as a deformation does not induce {\it boundary} non-zeros, the system will remain {\it integrable}.

\paragraph*{}Thus, only sensible and simplest way to make the system QI for sure is to deform $\sin u$ in $B$ and $C$ in a way that
it leads to non-zero boundary value, as the ${\cal O}\left(1/\lambda\right)$ charge remains unaffected, and thus
conserved still. This requirement is naturally satisfied by a {\it shift} deformation of the type,

\be
(B,~C)_{-1}\left(=-\frac{\delta}{4}\sin u\right)\rightarrow (B,~C)_{-1}+D_{-1}; \quad D_{-1}:=\sum_{m=1}\frac{r}{f_{m-1}}g_m(u),\label{N7}
\ee
where $g_m(u)$ are functions of $u$ and its derivatives, which is finite at the boundary. Then, for $m=n+1$, $n$ representing
the summation index in Eq. \ref{N4}, there will be a non-vanishing term on the RHS of Eq. \ref{N4}. For $m<n+1$, however,
$g_m$ will be multiplied with positive powers of $u_x$ and its derivatives, leading to conservation again. For $m>n+1$,
negative powers of $u_x$ and its derivatives will come into play, leading to infinities (non-conserved charges again).
One can very well truncate the vale of $m$ to avoid such infinities. All in all, quasi-integrability will be obtained
as desired.

\paragraph*{}One can very well attribute this deformation to the level of the Lagrangian (Eq. \ref{N3}) or the Hamiltonian
(Eq. \ref{N02}). As discussed in subsection \ref{1.1}, although the system dynamics contains higher derivatives, the canonical
variable is still $u$, making $\delta\sin u$ equivalent to the `potential' of the system. Therefore, the above deformation
of Eq. \ref{N7}, being QI, in principle corresponds to that of the sine-Gordon system \cite{FZ,FZ1}. However, the present 
deformation needs to be of shift nature (subsec. \ref{1.1}), unlike the power modification of the previous cases.

\subsection{Loop-algebraic treatment}
From Eq. \ref{N01} the Lax pair can be re-expressed as,

\be
L=-i\lambda\sigma_3+q\sigma_++r\sigma_- \quad{\rm and}\quad M=A\sigma_3+B\sigma_++C\sigma_-,\label{N8}
\ee  
where the Pauli matrices $\sigma_{3,\pm}$ satisfy $SU(2)$ algebra:

\be
\left[\sigma_3,\sigma_\pm\right]=\pm2\sigma_\pm \quad{\rm and}\quad \left[\sigma_+,\sigma_-\right]=\sigma_3,\label{N9}
\ee
allowing for the construction of the $SU(2)$ loop algebra \cite{N0},

\bea
&&\left[b^n,F_{1,2}^m\right]=2F_{2,1}^{m+n}, \quad \left[F_1^n,F_2^m\right]=\lambda b^{m+n}; \quad{\rm where}\quad,\nonumber\\
&&b^n=\lambda^n\sigma_3, \quad F_1^n=\frac{1}{\sqrt{2}}\lambda^n\left(\lambda\sigma_+-\sigma_-\right) \quad{\rm and}\quad F_2^n=\frac{1}{\sqrt{2}}\lambda^n\left(\lambda\sigma_++\sigma_-\right).\label{N10}
\eea
Such algebraic structure enables $sl(2)$ gauge rotation of the Lax pair, elegantly demonstrating the quasi-integrability 
of the system \cite{FZ}.
\paragraph*{}We now consider the QI deformation of Eq. \ref{N7}, modifying the curvature defined through the zero-curvature
condition in Eq. \ref{E3}; leading to,

\be
F_{tx}\rightarrow\bar{F}_{tx}=-\frac{1}{2}\bar{E}(u)\sigma_++\frac{1}{2}\bar{E}(u)\sigma_++{\cal X}\sigma_x,\quad\sigma_x=\sigma_++\sigma_-,\label{N11}
\ee
where $\bar{E}(u)$ is the Euler function yielding the Quasi-modified equation while equated to zero:

\be
u_{xt}+3\beta u_x^2u_{xx}+2\beta u_{xxxx}=\delta\sin u-4D_{-1},\label{N12}
\ee
and ${\cal X}$ is the {\it anomaly} term given as,

\be
{\cal X}=iD_{-1,x}.\label{N13}
\ee

\paragraph*{Gauge transformation:}Based on the standard gauge-fixing of the QI systems \cite{FZ,FZ1,N0}, we undertake a gauge transformation with respect
to the operator,

\be
g=\exp\left(\sum_{n=-\infty}^{\infty}J_n\right), \quad{\rm where},\quad J_n=a_1^nF_1^n+a_2^nF_2^n.\label{N14}
\ee 
The coefficients $a_{1,2}^n$ are to be chosen such that the new spatial Lax pair component $\bar{L}=gLg^{-1}+g_xg^{-1}$
contains only $b^n$s:

\be
\bar{L}\equiv\sum_n\beta_L^nb^n,\label{N15}
\ee 
making it diagonal in the $SU(2)$ basis. From the BCH formula:

$$e^XYe^{-X}=Y+[X,Y]+\frac{1}{2!}[X,[X,Y]]+\frac{1}{3!}[X,[X,[X,Y]]]+\cdots,$$
the gauge-transformed spatial component takes the form,

\be
\bar{L}=J_{n,x}+L+[J_n,L]+\frac{1}{2!}[J_m,[J_n,L]]+\frac{1}{3!}[J_l,[J_m,[J_n,L]]]+\cdots,\label{N16}
\ee
with summations understood over all semi-positive integers. Few of the lowest order commutators are,

\bea
\left[J_n,L\right]&=&2i\lambda\left(a_1^nF_2^n+a_2^nF_1^n\right)+\frac{q}{\sqrt{2}}\left(a_1^n-a_2^n\right)b^n+\frac{r}{\sqrt{2}}\left(a_1^n+a_2^n\right)b^{n+1},\nonumber\\
\frac{1}{2!}\left[J_m,\left[J_n,L\right]\right]&=&i\lambda\left(a_1^ma_1^n-a_2^ma_2^n\right)b^{m+n+1}-\frac{q}{\sqrt{2}}\left(a_1^n-a_2^n\right)\left(a_1^mF_2^{m+n}+a_2^mF_1^{m+n}\right)\nonumber\\
&&-\frac{r}{\sqrt{2}}\left(a_1^n+a_2^n\right)\left(a_1^mF_2^{m+n+1}+a_2^mF_1^{m+n+1}\right),\nonumber
\eea
\bea
\frac{1}{3!}\left[J_l,\left[J_m,\left[J_n,L\right]\right]\right]&=&-i\frac{2}{3}\lambda\left(a_1^ma_1^n-a_2^ma_2^n\right)\left(a_1^lF_2^{l+m+n+1}+a_2^lF_1^{l+m+n+1}\right)\nonumber\\
&&-\frac{q}{3\sqrt{2}}\left(a_1^n-a_2^n\right)\left(a_1^la_1^m-a_2^la_2^m\right)b^{l+m+n+1}\nonumber\\
&&-\frac{r}{3\sqrt{2}}\left(a_1^n+a_2^n\right)\left(a_1^la_1^m-a_2^la_2^m\right)b^{l+m+n+2},\nonumber\\
&&\vdots\label{N17}
\eea
The {\it gauge-fixing} condition of vanishing coefficients for $F_{1,2}^n$s leads to the order-by-order consistency relations:

\bea
&&{\cal O}\left(F_1^0\right):\quad a_{1,x}^0=\frac{r}{\sqrt{2}}+\frac{q}{\sqrt{2}}\left(a_1^0-a_2^0\right)a_2^0-2i\lambda a_2^0-\frac{q}{\sqrt{2}\lambda},\nonumber\\
&&{\cal O}\left(F_2^0\right):\quad a_{2,x}^0=-\frac{r}{\sqrt{2}}+\frac{q}{\sqrt{2}}\left(a_1^0-a_2^0\right)a_1^0-2i\lambda a_1^0-\frac{q}{\sqrt{2}\lambda},\nonumber\\
&&{\cal O}\left(F_1^1\right):\quad a_{1,x}^1=\frac{r}{\sqrt{2}}\left(a_1^0+a_2^0\right)a_2^0+\frac{q}{\sqrt{2}}\left[\left(a_1^0-a_2^0\right)a_2^1+\left(a_1^1-a_2^1\right)a_2^0\right]-2i\lambda a_2^1+\cdots,\nonumber\\
&&{\cal O}\left(F_2^1\right):\quad a_{2,x}^1=\frac{r}{\sqrt{2}}\left(a_1^0+a_2^0\right)a_1^0+\frac{q}{\sqrt{2}}\left[\left(a_1^0-a_2^0\right)a_1^1+\left(a_1^1-a_2^1\right)a_1^0\right]-2i\lambda a_1^1+\cdots,\nonumber\\
&&\qquad\qquad\qquad\quad\vdots\label{N18}
\eea
leaving behind the coefficients for $b^n$s as,

\bea
&&\beta_L^0=-i\lambda+\frac{q}{\sqrt{2}}\left(a_1^0-a_2^0\right),\nonumber\\
&&\beta_L^1=\frac{q}{\sqrt{2}}\left(a_1^1-a_2^1\right)+\frac{r}{\sqrt{2}}\left(a_1^0+a_2^0\right)+\left[i\lambda-\frac{q}{3\sqrt{2}}\left(a_1^0-a_2^0\right)\right]\left(a_1^0a_1^0-a_2^0a_2^0\right),\nonumber\\
&&\beta_L^2=\frac{q}{\sqrt{2}}\left(a_1^2-a_2^2\right)+\frac{r}{\sqrt{2}}\left(a_1^1+a_2^1\right)+2i\lambda\left(a_1^0a_1^1-a_2^0a_2^1\right)-\frac{q}{3\sqrt{2}}\Big[2\left(a_1^0-a_2^0\right)\left(a_1^0a_1^1-a_2^0a_2^1\right)\nonumber\\
&&\qquad\quad+\left(a_1^1-a_2^1\right)\left(a_1^0a_1^0-a_2^0a_2^0\right)\Big]-\frac{r}{3\sqrt{2}}\left(a_1^0+a_2^0\right)\left(a_1^0a_1^0-a_2^0a_2^0\right)+\cdots\nonumber\\
&&\qquad\vdots\label{N19}
\eea
Therefore, the transformed spatial Lax component $\bar{L}$ of Eq. \ref{N15} is completely determined \cite{FZ,FZ1,N0}. The
gauge-transformed temporal Lax component,

\be
\bar{M}=gMg^{-1}+g_tg^{-1}=\sum_n\left[\beta_M^nb^n+\varphi_1^nF_1^n+\varphi_2^nF_2^n\right],\label{N20}
\ee
can also be evaluated similarly. Some of the lowest order commutators are,

\bea
\left[J_n,M\right]&=&-2A\left(a_1^nF_2^n+a_2^nF_1^n\right)+\frac{B}{\sqrt{2}}\left(a_1^n-a_2^n\right)b^n+\frac{C}{\sqrt{2}}\left(a_1^n+a_2^n\right)b^{n+1},\nonumber\\
\frac{1}{2!}\left[J_m,\left[J_n,M\right]\right]&=&-A\left(a_1^ma_1^n-a_2^ma_2^n\right)b^{m+n+1}-\frac{B}{\sqrt{2}}\left(a_1^n-a_2^n\right)\left(a_1^mF_2^{m+n}+a_2^mF_1^{m+n}\right)\nonumber\\
&&-\frac{C}{\sqrt{2}}\left(a_1^n+a_2^n\right)\left(a_1^mF_2^{m+n+1}+a_2^mF_1^{m+n+1}\right),\nonumber\\
\frac{1}{3!}\left[J_l,\left[J_m,\left[J_n,M\right]\right]\right]&=&\frac{2}{3}A\left(a_1^ma_1^n-a_2^ma_2^n\right)\left(a_1^lF_2^{l+m+n+1}+a_2^lF_1^{l+m+n+1}\right)\nonumber\\
&&-\frac{B}{3\sqrt{2}}\left(a_1^n-a_2^n\right)\left(a_1^la_1^m-a_2^la_2^m\right)b^{l+m+n+1}\nonumber\\
&&-\frac{C}{3\sqrt{2}}\left(a_1^n+a_2^n\right)\left(a_1^la_1^m-a_2^la_2^m\right)b^{l+m+n+2},\nonumber\\
&&\vdots\label{N21}
\eea
that leads to the lowest order coefficients,

\bea
&&\beta_M^0=A+\frac{B}{\sqrt{2}}\left(a_1^0-a_2^0\right),\nonumber\\
&&\beta_M^1=\frac{B}{\sqrt{2}}\left(a_1^1-a_2^1\right)+\frac{C}{\sqrt{2}}\left(a_1^0+a_2^0\right)-\left[A+\frac{B}{3\sqrt{2}}\left(a_1^0-a_2^0\right)\right]\left(a_1^0a_1^0-a_2^0a_2^0\right),\nonumber\\
&&\beta_M^2=\frac{B}{\sqrt{2}}\left(a_1^2-a_2^2\right)+\frac{C}{\sqrt{2}}\left(a_1^1+a_2^1\right)-2A\left(a_1^0a_1^1-a_2^0a_2^1\right)-\frac{B}{3\sqrt{2}}\Big[2\left(a_1^0-a_2^0\right)\left(a_1^0a_1^1-a_2^0a_2^1\right)\nonumber\\
&&\qquad\quad+\left(a_1^1-a_2^1\right)\left(a_1^0a_1^0-a_2^0a_2^0\right)\Big]-\frac{C}{3\sqrt{2}}\left(a_1^0+a_2^0\right)\left(a_1^0a_1^0-a_2^0a_2^0\right)+\cdots\nonumber\\
&&\qquad\vdots\label{N22}
\eea

\bea
&&\varphi_1^0=a_{1,t}^0-\frac{C}{\sqrt{2}}-\frac{B}{\sqrt{2}}\left(a_1^0-a_2^0\right)a_2^0-2A a_2^0+\frac{B}{\sqrt{2}\lambda},\nonumber\\
&&\varphi_1^1=a_{1,t}^1-\frac{C}{\sqrt{2}}\left(a_1^0+a_2^0\right)a_2^0-\frac{B}{\sqrt{2}}\left[\left(a_1^0-a_2^0\right)a_2^1+\left(a_1^1-a_2^1\right)a_2^0\right]-2A a_2^1+\cdots,\nonumber\\
&&\qquad\vdots\label{N22a}
\eea

\bea
&&\varphi_2^0=a_{2,t}^0+\frac{C}{\sqrt{2}}-\frac{B}{\sqrt{2}}\left(a_1^0-a_2^0\right)a_1^0-2A a_1^0-\frac{B}{\sqrt{2}\lambda},\nonumber\\
&&\varphi_2^1=a_{2,t}^1-\frac{C}{\sqrt{2}}\left(a_1^0+a_2^0\right)a_1^0+\frac{B}{\sqrt{2}}\left[\left(a_1^0-a_2^0\right)a_1^1+\left(a_1^1-a_2^1\right)a_1^0\right]-2A a_1^1+\cdots,\nonumber\\
&&\qquad\vdots\label{N22b}
\eea
\paragraph*{The transformed curvature:}The above gauge transformation further yields the curvature as,

\be
\bar{F}_{tx}=gF_{tx}g^{-1}\equiv{\cal X}\sum_n\left(f_0^nb^n+f_1^nF_1^n+f_2^nF_2^n\right),\label{N23}
\ee
with the lowest order commutators,

\bea
\left[J_n,F_{tx}\right]&=&\frac{{\cal X}}{\sqrt{2}}\left[\left(a_1^n-a_2^n\right)b^n+\left(a_1^n+a_2^n\right)b^{n+1}\right],\nonumber\\
\frac{1}{2!}\left[J_m,\left[J_n,F_{tx}\right]\right]&=&-\frac{{\cal X}}{\sqrt{2}}\Big[\left(a_1^n-a_2^n\right)\left(a_1^mF_2^{m+n}+a_2^mF_1^{m+n}\right)\nonumber\\
&&\qquad\quad+\left(a_1^n+a_2^n\right)\left(a_1^mF_2^{m+n+1}+a_2^mF_1^{m+n+1}\right)\Big],\nonumber\\
\frac{1}{3!}\left[J_l,\left[J_m,\left[J_n,F_{tx}\right]\right]\right]&=&-\frac{{\cal X}}{3\sqrt{2}}\left(a_1^la_1^m-a_2^la_2^m\right)\left[\left(a_1^n-a_2^n\right)b^{l+m+n+1}+\left(a_1^n+a_2^n\right)b^{l+m+n+2}\right],\nonumber\\
&&\vdots\label{N24}
\eea
It is to be noted that the rotated curvature obtained above is on-shell, {\it i. e.}, owing to the gauge-fixing above, 
Eq. \ref{N12} can be applied. Therefore, it depends only on the anomaly function ${\cal X}$ of the QI deformation, and 
duly vanishes for ${\cal X}=0$. A few lowest order co-efficients of the curvature are expressed below:

\bea
&&f_0^0=\frac{1}{\sqrt{2}}\left(a_1^0-a_2^0\right),\nonumber\\
&&f_0^1=\frac{1}{\sqrt{2}}\left(a_1^1-a_2^1\right)+\frac{1}{\sqrt{2}}\left(a_1^0+a_2^0\right)-\frac{1}{3\sqrt{2}}\left(a_1^0-a_2^0\right)\left(a_1^0a_1^0-a_2^0a_2^0\right),\nonumber\\
&&f_0^2=\frac{1}{\sqrt{2}}\left(a_1^2-a_2^2\right)+\frac{1}{\sqrt{2}}\left(a_1^1+a_2^1\right)-\frac{1}{3\sqrt{2}}\Big[2\left(a_1^0-a_2^0\right)\left(a_1^0a_1^1-a_2^0a_2^1\right)\nonumber\\
&&\qquad\qquad+\left(a_1^0a_1^0-a_2^0a_2^0\right)\left\{\left(a_1^1-a_2^1\right)+\left(a_1^0+a_2^0\right)\right\}\Big]+\cdots\nonumber\\
&&\qquad\vdots\label{N25a}
\eea

\bea
&&f_{1,2}^0=\left(\lambda^{-1}\mp 1\right)-\frac{1}{\sqrt{2}}\left(a_1^0-a_2^0\right)a_{2,1}^0,\nonumber\\
&&f_{1,2}^1=-\frac{1}{\sqrt{2}}\left[\left(a_1^0-a_2^0\right)a_{2,1}^1+\left(a_1^1-a_2^1\right)a_{2,1}^0+\left(a_1^0+a_2^0\right)a_{2,1}^0\right],\nonumber\\
&&f_{1,2}^2=-\frac{1}{\sqrt{2}}\left[\left(a_1^1-a_2^1\right)a_{2,1}^1+\left(a_1^0+a_2^0\right)a_{2,1}^1+\left(a_1^1+a_2^1\right)a_{2,1}^0\right]+\cdots\nonumber\\
&&\qquad\vdots\label{N25b}
\eea 
Finally, from the definition $\bar{F}_{tx}=\bar{L}_t-\bar{M}_x+[\bar{L} , \bar{M}]$ of the rotated curvature, by
substituting the coefficients obtained above, one finds the consistency conditions:

\bea
&&\beta_{L,t}^n-\beta_{M,x}^n={\cal X}f_0^n \quad{\rm and}\quad \varphi_{-,x}^n=-{\cal X}f_-^n-2\beta_L^n\sum_m\varphi_-^m\lambda^m,\label{N26}\\
&&{\rm where},\quad \varphi_-^n:=\varphi_1^n-\varphi_-2^n \quad{\rm and}\quad f_-^n:=f_1^n-f_2^n,\nonumber
\eea
entirely evaluate all the coefficients with the aid of the gauge-fixing conditions in Eq.s \ref{N18} \cite{N0}.

\paragraph*{Quasi-conservation:}Following the treatment in Ref.s \cite{Z,FZ1,N0}, the {\it lowest order} quasi-conserved charge
is, 

\be
Q^0:=\int_x\beta_L^0\equiv\int_x \left(-i\lambda+\frac{1}{\sqrt{2}}qa_-^0\right);\quad a_-^0:=a_1^0-a_2^0,\label{N27}
\ee
which is expected to be conserved for the system to be QI. This is consistent with the definition of the continuity
expression in Eq. \ref{N5}. From the lowest order ($n=0$) contribution of first of the Eq.s \ref{N26}, 

\be
\frac{dQ^0}{dt}=\frac{1}{\sqrt{2}}\int_x \left(qa_-^0\right)_t\equiv\frac{1}{\sqrt{2}}\int_x{\cal X}a_-^0=\frac{i}{\sqrt{2}}\int_xD_{-1,x}a_-^0,\label{N28}
\ee
modulo vanishing total derivatives of functions of $u$, which is sensibly assumed to vanish asymptotically. For a given
$u_x$, $a_-^0$ is uniquely determined from the gauge-fixing and consistency conditions. Then, it is always possible to
choose the deformation functions as,

\be
g_m(u)=\frac{f_{m-1}}{r}\left(a_-^0\right)^p,\quad p>0,\label{N28}
\ee
for {\it all} $m=1,2,3,\cdots$, leading to,

\be
\frac{dQ^0}{dt}=\frac{i}{\sqrt{2}}\frac{\sum_m}{1+p}\int_x\left(a_-^0\right)_x^{1+p}\equiv 0,\label{N29}
\ee
leading to a conserved charge all the time. This ensures at least {\it one} conserved charge for the deformed system, and
thus, quasi-integrability. This has been obtained independent to the more comprehensive direct observation of the 
previous section without considering the loop algebraic structure.

\section{Conclusions}

We have considered the combination of two well known dynamical systems, namely,
Frenkel-Kontorova and Fermi-Pasta-Ulam models. This new dynamical system, proposed by Kudryashov,
has been obtained by taking the continuum limit approximation for $N \to \infty$ and
$h \to 0$. This continuous equation becomes the mixture of {\it generalized potential KdV equation}
and the {\it sine-Gordon equation}. Using Wadati-Konno formalism we have studied the B\"acklund transformation
from Riccati form of inverse method. In the second half of the paper we have studied the quasi-integrable deformation
of this new equation. Earlier we have shown that the quasi-integrable deformation of the KdV system \cite{N0} is indeed
possible, provided the loop-algebraic generalization has been considered, and in fact we also know the quasi-integrable
deformation of the sine-Gordon \cite{FZ,FZ1} and super sine-Gordon equations \cite{N1}. Here, in succession, we have
studied quasi-integrable deformation of the mixed generalized potential KdV and sine-Gordon equation.

\section*{Acknowledgement} The authors
are grateful to Professors Luiz. A. Ferreira, Wojtek J. Zakrzewski and Betti Hartmann for their encouragement,
various useful discussions and critical reading of the draft.


\begin{thebibliography}{99}

\bibitem{Kud1}N A Kudryashov, \textit{Integrable model of nonlinear dislocations}, arXiv:1611.06813v1[nlin.SI]

\bibitem{KF} T.A. Kontorova, Ya. I. Frenkel, \textit{On theory of plastic deformation}, 8
(1938) JETP, 89, 1340, 1349 (in Russian)

\bibitem{BK} Oleg M. Braun, Yuri S. Kivshar, \textit{Nonlinear dynamics of the Frenkel-
Kontorova model}, Physics Reports 306 (1998) 1-108.

\bibitem{FPU} E. Fermi, J.R. Pasta, S.Ulam, \textit{Studies of nonlinear problems}, Report
LA-1940, 1955. Los Alamos: Los Alamos Scientific Laboratory.

\bibitem{FPU1} M.A. Porter, N.J. Zabusky, B. Hu, D.K. Campbell, \textit{Fermi, Pasta,
Ulam and the Birth of Experimental Mathematics}, American Scientist,
97(3)(2009), 214-221. doi:10.1511/2009.78.214.

\bibitem{ZK} N.J. Zabusky, M.D. Kruskal, \textit{Interactions of solitons in a collisionless
plasma and the recurrence of initial states}, Physical Review Letters,
15(6),(1965) 240-243.

\bibitem{FT} L.D. Faddeev and L.A. Takhtajan, \textit{Hamiltonian Methods in the Theory of Solitons}. Berlin: Springer-Verlag, 1987.

\bibitem{Das} A. Das, \textit{ Integrable models}, World Scientific, Singapore (1989).

\bibitem{KW} K. Konno and M. Wadati, \textit{Simple Derivation of B\"acklund Transformation
from Riccati Form of Inverse Method}, Progress of Theoretical Physics
Volume 53, Issue 6, 1652-1656.

\bibitem{KW1} M. Wadati, H. Sanuki and K. Konno, \textit{ Relationships among Inverse Method, B\"acklund Transformation and an 
Infinite Number of Conservation Laws}, Progress of Theoretical Physics
Volume 53, Issue 2, 419-436.

\bibitem{FerrSG} S. Ferreira, L. Girardello and S. Sciuto, \textit{An infinite set of conservation laws of the supersymmetric sine-gordon theory}, Phys. Lett. B, {\bf 76} (1978) 303.

\bibitem{FZ} L. A. Ferreira and W. J. Zakrzewski, \textit{ The concept of quasi-integrability:  a concrete example}, JHEP, {\bf 05}, 130 (2011).
\bibitem{FZ1} L. A. Ferreira, G. Luchini and W. J. Zakrzewski, \textit{The  Concept  of  Quasi-Integrability}, Nonlinear and Modern Mathematical Physics AIP Conf. Proc., {\bf 1562}, 43 (2013).

\bibitem{N0}K. Abhinav and P. Guha, \textit{Quasi-Integrability of The KdV System}, arXiv:1612.07499 [math-ph].

\bibitem{N1}K. Abhinav and P. Guha, \textit{ Quasi-Integrability in Supersymmetric Sine-Gordon Models}, EPL 116 (2016) 10004.


\end{thebibliography}
\end{document}